\pdfoutput=1
\documentclass{JINST}
\let\ifpdf\relax

\usepackage{amsmath}   
\usepackage{amssymb}
\usepackage{graphicx}        
\usepackage{subfigure}          
\usepackage{array}
\usepackage{ulem}

\title{Signal and noise of Diamond Pixel Detectors at High Radiation Fluences}

\author{Jieh-Wen Tsung$^a$, Miroslav Havranek$^a$, Fabian H{\"u}gging$^a$, Harris Kagan$^b$, Hans Kr{\"u}ger$^a$, and Norbert Wermes$^a$\thanks{Corresponding author}\\
\llap{$^a$}Bonn University, \\
Physikalisches Institut, Nussallee 12, D-53115 Bonn, Germany\\
\llap{$^b$}The Ohio State University, \\
Department of Physics, 191 W. Woodruff Ave., Columbus, Ohio 43210, U.S.A
\thanks{Work supported by the German Ministerium f{\"u}r Bildung, Wissenschaft, Forschung und Technologie (BMBF) under contract no. 05H09PD2 and by US Department of Energy grant DE-FG02-91ER40690.}
\\
E-mail: \email{wermes@uni-bonn.de}
}

\abstract{CVD diamond is an attractive material option for LHC vertex detectors mainly because of its strong radiation-hardness causal to its large band gap and strong lattice. In particular, pixel detectors operating close to the interaction point profit from tiny leakage currents and small pixel capacitances of diamond resulting in low noise figures 
when compared to silicon. On the other hand, the charge signal from traversing high energy particles is smaller in diamond than in silicon by a factor of about 2.2. Therefore, a quantitative determination of the signal-to-noise ratio (S/N) of diamond in comparison with silicon at fluences in excess of 10$^{15}$ n$_{eq}$ cm$^{-2}$, which are expected for the LHC upgrade, is important. 
Based on measurements of irradiated diamond sensors and the FE-I4 pixel readout chip design and performance,
we determine the signal and the noise of diamond pixel detectors irradiated with high particle fluences. 
To characterize the effect of the radiation damage on the materials and the signal decrease,  
the change of the mean free path $\lambda_{e/h}$ of the charge carriers is determined as a function of irradiation fluence.
We make use of the FE-I4 pixel chip developed for ATLAS upgrades to realistically estimate the expected noise figures: the expected leakage current at a given fluence is taken from calibrated calculations and the pixel capacitance is measured using a purposely developed chip (PixCap). We compare the resulting S/N figures with those for planar silicon pixel detectors using
published charge loss measurements and the same extrapolation methods as for diamond. 
It is shown that the expected S/N of a diamond pixel detector with pixel pitches typical for LHC, exceeds that of planar silicon pixels at fluences beyond 10$^{15}$ particles cm$^{-2}$, the exact value only depending on the maximum operation voltage assumed for irradiated silicon pixel detectors.
}


\begin{document}
\section{Introduction}\label{sec:intro}
After almost three decades of successful operation of gaseous or silicon microstrip vertex detectors at most high energy physics collider experiments, the LHC detectors ATLAS~\cite{ATLAS_detector_paper_2008}, CMS~\cite{CMS_detector_paper_2008}, and ALICE~\cite{ALICE_detector_paper_2008} have for the first time installed large silicon pixel detectors~\cite{ATLAS_pixel_paper_2008,CMS_pixel_paper_2008,Santoro:2009zza} as their innermost detection elements for precision tracking and vertexing close to the collision point. Pixel detectors are capable of coping with
the large data rates at the LHC and can survive the large radiation fluences in the vicinity close to the collision point. 
Since the start of the LHC in 2009 they have proven to be precise and reliable detection
devices~\cite{hirsch_pixel2010,dellasta_pixel2010,bean_pixel2010,riedler_pixel2010,langenegger_pixel2010}.
In the context of the LHC upgrade program~\cite{HL-LHC2011}, in particular the expected increase in luminosity by up to a factor 10 compared to the present LHC design~\cite{LHC-machine} and the corresponding increase in particle rates and radiation levels, strong constraints on detector
layouts and materials are imposed. While steady progress in microelectronics helps to cope with the resulting data rates, 
new sensor solutions must be targeted to withstand particle fluences in excess of 10$^{16}$ cm$^{-2}$ absorbed over the machine's lifetime.

In attempting to cope with this fierce radiation environment, sensor materials with a strong intrinsic radiation resistance come to mind. In this category diamond, grown by chemical vapor deposition (CVD),
is a candidate due to its large band gap of 5.5 eV and its strong lattice, for which a minimum displacement energy of 43 eV~\cite{Koike} is needed to
kick-off an atom from the lattice (25 eV in Si~\cite{Lintbook1980}). 
CVD diamond detectors are produced as poly-crystalline (pCVD) and single-crystal (scCVD) detectors.
The large band gap also results in quasi negligible leakage current, an important noise factor. Moreover, the pixel capacitance should be significantly smaller for diamond than for silicon pixels, not only because of the smaller dielectric constant ($\epsilon_{CVD}$= 5.7; $\epsilon_{Si}$= 11.9) but also because of the more complex structure of implants and guard rings in the case of silicon~\cite{planar-pixels}. On the negative
side, also due to the larger band gap, the average energy required to liberate an e/h-pair in diamond is 13.1~eV, 3.6 times larger than for silicon, resulting in a smaller signal, at least before irradiation. 
To assess whether this still holds true after heavy irradiation, the development of signal and signal-to-noise (SNR) as a function of radiation fluence is determined in this paper.
As a model pixel system we adopt the current ATLAS pixel modules developed for the insertable B-layer project, IBL~\cite{hugging_pixel2010} with silicon pixel sensors and its diamond beam monitor companion (DBM).
Both are equipped with the pixel readout chip FE-I4 ~\cite{FE-I4_MB2009,FE-I4_MGS2011}.
While the detection efficiency and the fake hit probability also depends on the chip's threshold, 
SNR has been chosen as the characterizing quantity in order to avoid assuming a concrete threshold setting which depends on
a variety of tuning and environment parameters (especially at high fluences) which are difficult to predict. It is straight forward, however,  to determine the detection efficiency and/or fake hit probability for any given threshold once SNR is given.      

First, we describe the signal and its development with increasing particle fluence. The damage is characterized in terms of the mean free path $\lambda_{e/h}$ of the charge carriers: 
\begin{equation}\label{eq:mfp}
\lambda_{e/h} = {\rm v}_{e} \tau_{e} + {\rm v}_{h} \tau_{h} \ ,
\end{equation}
where v$_{e/h}$ and $\tau_{e/h}$ are the velocities and lifetimes of the charge carriers. 

Then we determine the expected noise using the ATLAS pixel readout chip FE-I4~\cite{FE-I4_MB2009, FE-I4_MGS2011} as a reference for calculations and simulations.
Two important noise parameters, the  leakage current (i$_{leak}$) and the input capacitance coupled to the preamplifier (C$_D$), are either calculated or measured, respectively.
Finally, the SNR as a function of fluence is determined from these ingredients. In these steps we keep the corresponding results for planar silicon pixel sensors~\cite{planar-pixels} operated with the same readout chip and biased with voltages up to 600 V as a reference. 

\section{Diamond sensor characterization}\label{sec:exp_basis}
We evaluate the signal to noise of diamond pixel sensors that are bonded to the ATLAS pixel readout chip FE-I4~\cite{FE-I4_MGS2011}.
This chip is the successor of the FE-I3 chip~\cite{Peric2006} currently installed in ATLAS, and has been 
developed to cope with the demands of the next generation of pixel detectors. 
The FE-I4 chip has been bonded to different sensor types, among them planar~\cite{planar-pixels} and 3D-silicon~\cite{3D-Si} sensors as well as diamond sensors~\cite{diamond_RD42}. 
The area of one pixel is 50 $\times$ 250 $\rm \mu$m, and the entire chip has 26880 pixels in 80 columns and 336 rows.
The FE-I4 will serve as the base readout chip for the pixel detector upgrade developments in the coming years and is this particularly suited for this study.

\begin{figure}
\begin{center}
	\includegraphics[width=0.9\textwidth]{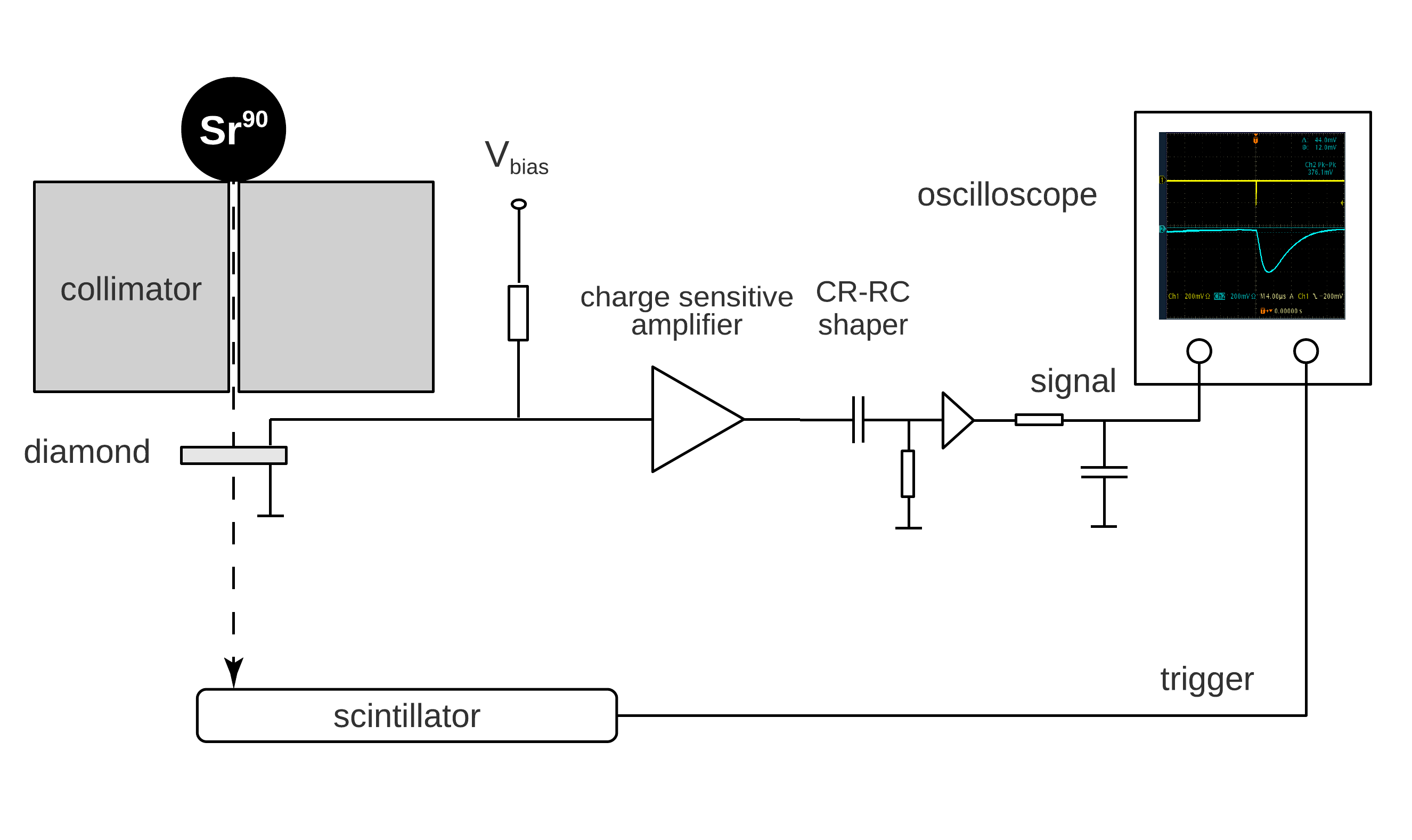}
\caption[]{Measurement setup to characterize the diamond sensor samples using a $^{90}$Sr source. 	\label{fig:Sr90test}} 
\end{center}
\end{figure}

The measurement of the change in signal charge before and after irradiation should not depend on the type of readout used
and is thus performed using detectors equipped with a single channel charge sensitive readout connected to 
a readout pad on the diamond sensor. 
Because the effects of radiation damage differ largely for different energies, the scCVD and pCVD diamond pad detectors 
have been irradiated with 24 GeV protons at the CERN SPS~\cite{Glaser24GeVPS} and with 25 MeV protons at the Karlsruhe Synchrotron~\cite{Dierlamm25MeV} to various fluences. Their charge signal has been measured before and after
the irradiation to assess the relative signal loss.
The measurement is done with the setup shown in Fig.~\ref{fig:Sr90test}.  A $^{90}$Sr $\beta$-source with a 2.28 MeV endpoint energy is used. In order to select only straight, high energetic tracks of the $^{90}$Sr spectrum that closely represent the energy deposit of a minimum ionizing particle, electrons need to pass a collimator and reach a scintillator for triggering. Range calculations show that with the given material the electrons must have a minimum energy of about 1.05 MeV corresponding to $\beta\gamma \approx 2$ to trigger and are thus considered very close to minimum  ionizing~\cite{PDG}. Hardening the $\beta$-spectrum by adding additional material sheets into the electron's flight path between sensor and scintillator did not change the measured average signal. 
The charge deposited by the triggered electrons in the sensor is read out by a charge sensitive preamplifier~\cite{CSA-preamp} and a pulse shaper system\cite{Kubota:1991ww} with a 2.5 $\mu$s shaping time. 
The analysis of the data is shown in section ~\ref{sec:damage}.

To compare the noise calculations/simulations (see section~\ref{sec:noise}) with the noise of real devices a few sensor - FE-I4 pixel modules have been assembled. The noise of these modules is measured
by charge injections to the front-end and recording so-called threshold S-curves ~\cite{S-curve}.

\section{The signal}\label{sec:signal}
\subsection{Energy deposit and charge signal}\label{sec:dEdx}
The absolute signal charge is calculated from the energy deposit in the sensor and the average energy needed to create an electron/hole pair (see section~\ref{sec:intro}).  
The energy released per cm in the diamond or silicon sensors is given in \cite{PDG} including the effects of $\delta$-electrons leaving the sensor.
The most probable value is calculated following~\cite{PDG} 
\begin{equation}\label{eq:Landau-Vavilov-Bichsel}   
MPV  = \xi 
\left [ \ln \frac{2 m_e c^2 \beta^2 \gamma^2 }{I}
+ \ln \frac{\xi }{I} + 0.2 - \beta^2 - \delta \right ]
\end{equation}
where $\xi = \frac{K}{2} \frac{Z}{A} \frac{x}{\beta^2}$ MeV, for a detector with a thickness of x g cm$^{-2}$;
$\delta$ parametrizes the density effect on the
energy loss and $I$ is the average ionization energy. The respective values for diamond and silicon are given in 
Table~\ref{tab:par_bethe}.

\begin{table}[h]
\begin{center}
\begin{tabular}{lll}
  \hline
  parameter      & diamond        & Si      \\
  \hline
  I              	& 81 eV              & 174 eV         \\
  $\delta$       	& 1.84 	           & 0.95           \\
  $w_{i}$       	& 13.1 eV           & 3.61 eV     \\
  \hline
\end{tabular}
\caption{Energy loss parameters for 
equation (3.1) and e/h creation energy $w_i$ for diamond and silicon.  
$\delta$ is taken from ref.~\cite{Sternheimer:1971zz} for 1 GeV pions ($\beta\gamma = 7.2$). 
} \label{tab:par_bethe}
\end{center}
\end{table}

\begin{figure}[thb]
\begin{center}
\includegraphics[height=0.4\textheight]{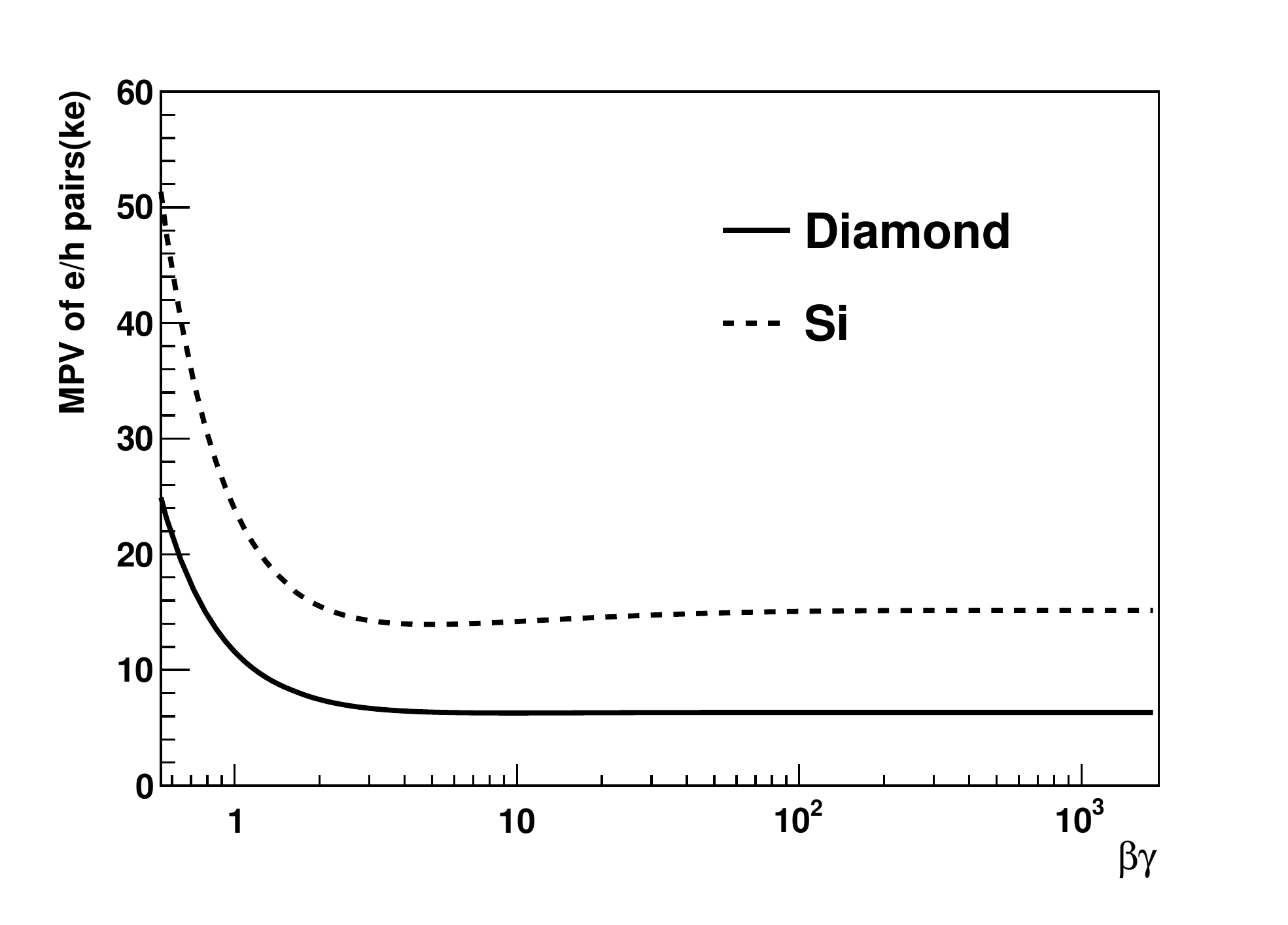}
\caption[]{Generated signal charge (e/h pairs, MPV) in a 200$\mu$m diamond or silicon sensor before irradiation.}
\label{fig:signal}
\end{center}
\end{figure}

The energy deposit in diamond per unit thickness is larger than in silicon mainly due to the larger density $\rho$ (see Fig.~\ref{fig:signal}). The ratio for a 1 GeV incident pion ($\beta\gamma$ = 7.2)  is
\begin{eqnarray}\label{eq:ratio}
\frac{MPV_{\rm{diamond}}} {MPV_{\rm{Si}}} & \approx & 1.63       \ .
\end{eqnarray}

The signal charge released for a given deposited energy is obtained by dividing the energy deposition (MPV) by the energy required to generate an e/h-pair $w_i$ (Table~\ref{tab:par_bethe}). For a 200$\mu$m thick sensor,  this yields about 14\,000 (Si) and 6\,300 (diamond) e/h-pairs, respectively for 1 GeV pions before irradiation.
%

\subsection{Radiation damage and signal losses}\label{sec:damage}
If the mean free path of the charge carriers $\lambda_{e/h}$ is much larger than the thickness of the sensor, the signal measured by a charge sensitive amplifier corresponds to the collection of the entire generated charge. Heavy irradiation introduces trapping centers into the sensor crystal which in turn reduces $\lambda_{e/h}$. 
Especially in silicon, additional effects like the change of the effective doping concentration, and hence the internal electric field, as well as increased leakage current generation also occur, which are treated in section~\ref{sec:noise}. 

\begin{figure}
\begin{center}
\includegraphics[width=0.8\textwidth]{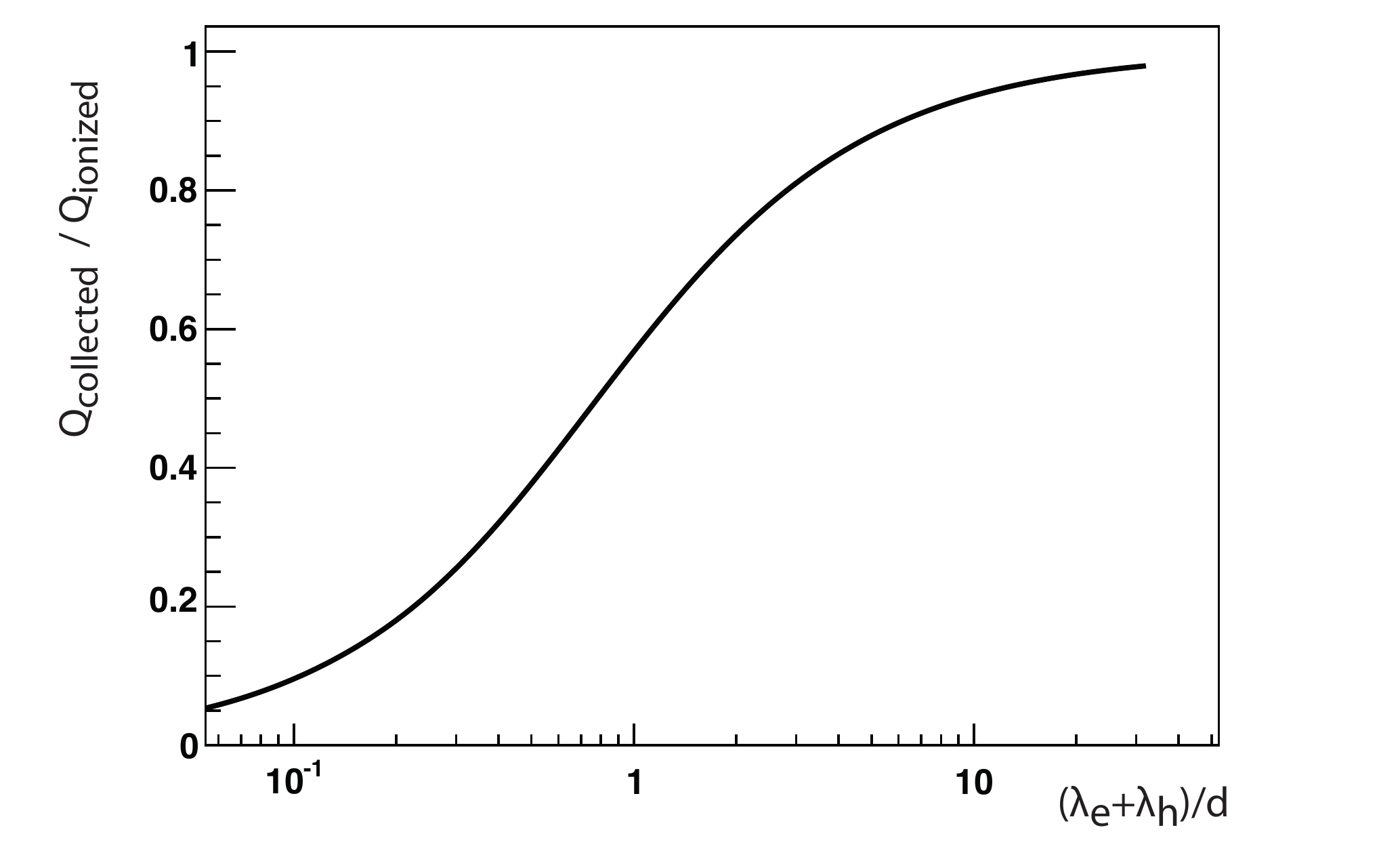}
\caption[]{Relation between charge collection distance CCD and carrier mean free path 
$\lambda_{e/h} = \lambda_e + \lambda_h$, both normalized to the sensor thickness d.}
\label{fig:mfp}
\end{center}
\end{figure}

\paragraph*{Radiation effects in diamond}
In former studies of diamond sensors~\cite{RD42-LHCC-2007,RD42-LHCC-2008}, the signal loss in diamond due to irradiation has been characterized by the charge collection distance CCD which is defined as the ratio of the measured collected charge and the charge generated by ionization multiplied by the detector thickness:
\begin{eqnarray}
CCD = \frac{Q_{collected}}{Q_{ionized}} \times d
\end{eqnarray}
The charge collection distance ceases to be a good characterization quantity when $\lambda_{e/h}$ is larger than the sensor thickness which is generally the
case for unirradiated silicon but also for present day single crystal diamond sensors~\cite{large_CCD}. In general, $\lambda_{e/h}$ itself  is the appropriate quantity to characterize the signal loss. It relates to CCD as~\cite{Zhao1994}:
\begin{equation}\label{eq:mfp_to_CCD}
\frac{CCD}{d} = \frac{Q_{collected}}{Q_{ionized}} = \frac{\lambda_{e/h}}{d} \cdot \left[ 1-\frac{\lambda_{e/h}}{d} \left ( 1-e^{-\frac{d}{\lambda_{e/h}}} \right ) \right ] \quad + \quad \left( e \leftrightarrow h \right)  .
\end{equation}
The relation is displayed in Fig.~\ref{fig:mfp}. For large $\lambda_{e/h}/d$, CCD reaches $d$.
For $\lambda_{e/h}/d \ll 1$ CCD and $\lambda_{e/h}$ are about the same.

\paragraph*{Radiation effects in silicon}
While for diamond the effect of strong irradiation causes a loss of signal charge due to trapping and a reduction of $\lambda_{e/h}$, 
for silicon additional effects must be taken into account.
Radiation damage effect to silicon have been studied intensively and are reliably described by the so called NIEL 
hypothesis~\cite{Lindstroem} , according to which all damage can be traced to point and cluster defects created by non-ionizing energy loss in the lattice.  
With increasing radiation the effective doping concentration changes 
leading to a smaller depletion depth at the same bias voltage. 
At very high fluences, and with high electric fields inside the sensor, charge multiplication effects have been 
reported~\cite{Affolder_25MeV}. 
Finally, also the geometry of the pixel electrodes plays an important role. Conventional sensors with planar pixel electrodes
require high operating voltages after irradiation, and cannot be fully depleted. In contrast, 
3D-silicon pixels~\cite{3D-Si} remain fully depleted even at moderate bias voltages. 
While these effects influence the signal, the noise is dominated by the increase in leakage current due to irradiation, common to both planar and 3D pixel sensors. 3D sensors also have more capacitance than planar sensors (see section~\ref{sec:pixcap_measurements}). 

\paragraph*{Comparison of diamond with silicon}
These various facts render a direct comparison of diamond with silicon signals difficult. We have thus chosen to compare 
our measurements obtained using diamond sensors with published silicon data~\cite{Affolder_25MeV,Affolder_24GeV} 
for about the same electric field strength ($\sim$1.8-2.2 V/$\mu$m) before irradiation. 
This corresponds to 900 V on 500 $\mu$m thick diamond sensors and 600 V for 310 $\mu$m thick silicon sensors when 
operated in over-depletion~\cite{Affolder_25MeV,Affolder_24GeV}.
This choice corresponds also to roughly the maximum bias voltage supplied at LHC pixel detectors. Much higher field strengths
also are more likely to show effects of charge multiplication after high irradiation. This would make a comparison between sensors more difficult. 

At LHC, a mixture of particles with a broad energy spectrum damages the detectors and electronics with energy and particle type dependent damage effects. Low energy (MeV) particles create larger damage than high energy (GeV) ones.
Irradiations of scCVD and pCVD diamond sensors have been performed by the RD42 collaboration
~\cite{diamond_RD42} for protons and neutrons at various energies between 25 MeV and 24 GeV. 
In the absence of a generally accepted NIEL damage normalization for diamond, e.g. to 1 MeV neutron damage, we refer in this paper to these two energies for proton damage because (a) they have been characterized by ourselves
and (b) they correspond well to the high and low energy parts of the particle energy spectrum at the LHC.
For silicon we use published data of irradiated Si planar n-in-p FZ pixel sensors~\cite{Affolder_25MeV,Affolder_24GeV}. 
The sensors have a thickness of 310$\mu$m and are read out by strip electrodes with 80$\mu$m pitch.
The measurements were performed with a bias voltage of 500 V and 700 V, respectively, and have been averaged for an assumed
bias voltage of 600 V.

$\lambda_{e/h}$ is obtained by using eq.~\eqref{eq:mfp_to_CCD}, where the collected charge $Q_{collected}$ is by measurement (diamond) or from publication (Si), and the generated charge $Q_{ionized}$ is calculated from eq.~\eqref{eq:Landau-Vavilov-Bichsel}.
\begin{figure}
\begin{center}
\subfigure[25 MeV protons]{
    \includegraphics[width=0.45\textwidth]{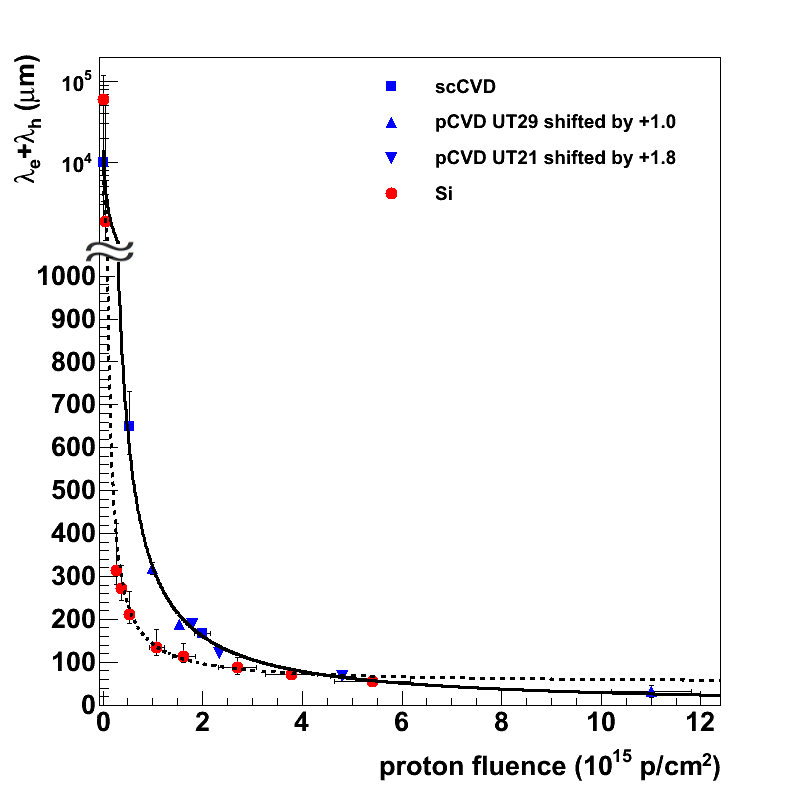}
    \label{fig:25MeVp_damage}}
    \hskip 0.5 cm
\subfigure[24 GeV protons]{
    \includegraphics[width=0.45\textwidth]{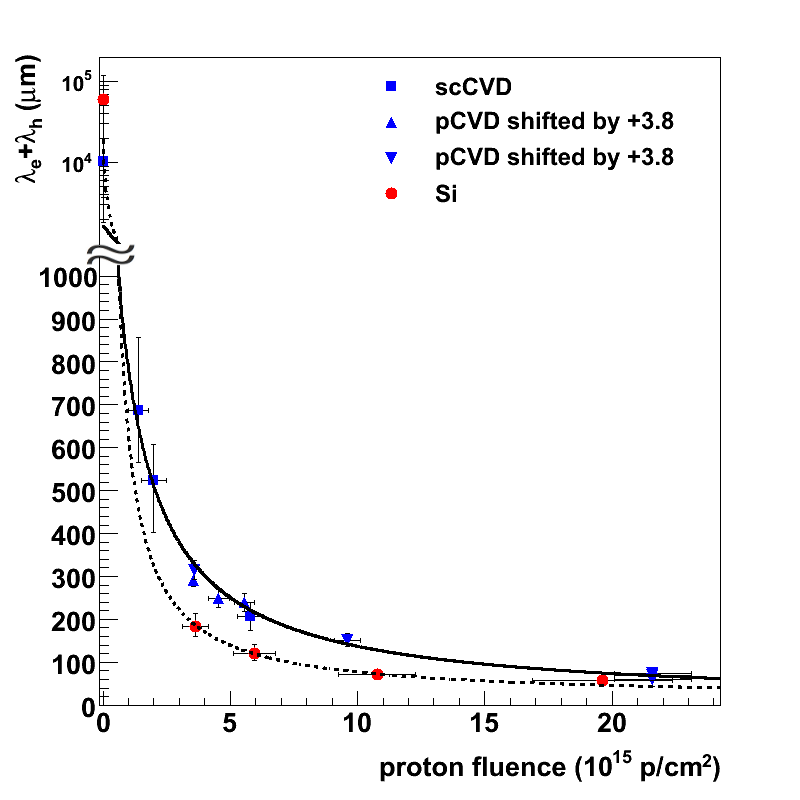}
    \label{fig:25GeVp_damage}}
\caption[]{Damage curves for 25MeV (a) and 24 GeV (b) proton irradiation on diamond and silicon. The diamond data contain measurements using
single crystal (scCVD) and poly-crystalline (pCVD) samples which are shifted to the right by the amount of fluence indicated in the legend. The silicon data are extracted from~\cite{Affolder_25MeV, Affolder_24GeV}. Note that the y-axis is cut above 1000 in order to display the data at low fluences.} 
\label{fig:damage_curves}
\end{center}
\end{figure}
Figure~\ref{fig:damage_curves} shows damage curves measured for irradiations with 25 MeV and 24 GeV protons, respectively. Displayed is $\lambda_{e/h}$ as a function of the radiation fluence. The shape of the damage curves generally follows the parametrization~\cite{Zhao1994}:
\begin{eqnarray} \label{eq:damage_curve}
\frac{1}{\lambda_{e/h}}= \frac{1}{\lambda_{0}} + k\,\Phi   \qquad \Leftrightarrow  \qquad
\lambda_{e/h}= \frac{\lambda_{0}}{1+\lambda_{0} k \Phi}
\end{eqnarray}
where $\Phi$ is the fluence and $\lambda_0$ is the mean free path before irradiation. $k$ is the damage constant 
parametrizing the material's irradiation hardness. Fig.~\ref{fig:damage_curves} contains data for diamond in mono-crystalline  (single crystal, scCVD) and poly-crystalline (pCVD) form. Note that the pCVD
diamond damage curves follow the same shape as the scCVD, but must be shifted to the right by a sample-dependent amount to fit the same curve.
This means that their initial mean free path $\lambda_0$ is shorter than that of the scCVD sample, because their less homogeneous substrate volume contains grain structures leading to a larger trapping center density~\cite{lari2005}.
Note however, that the shift is comparatively small once fluences above 10$^{15}$ cm$^{-2}$ have been reached.  A fit to the functional form of eq.~\eqref{eq:damage_curve} is performed to extract the damage constant $k$. In this the pCVD data shift 
value is determined yielding minimal overall $\chi^2$.

For silicon sensors damaged from radiation fluences above 10$^{15}$ n$_{eq}$ cm$^{-2}$ and operated at high bias voltages, 
charge multiplication effects have been reported~\cite{Affolder_25MeV}. As a consequence, the silicon data in 
Fig.~\ref{fig:damage_curves} do not follow the functional form of eq.~\ref{eq:damage_curve}.  In order to account for possible  
charge multiplication effects the parametrization of eq.~\ref{eq:damage_curve} has been changed by an additional term:
\begin{eqnarray} \label{eq:damage_curve_2}
\lambda_{e/h}= \frac{\lambda_{0}}{1+\lambda_{0} k \Phi} + \alpha \Phi^{\beta}
\end{eqnarray}
\sloppy
The corresponding fits yield $\beta \approx 0$ for all diamond and Si data and $\alpha = (51 \pm 10)\,\mu$m and 
$(14 \pm 19)\,\mu$m for silicon damaged by 25 MeV and 24 GeV protons, respectively. This supports the 
observation of an onset of charge multiplication in the low energy radiation data (25 MeV) which is more damaging than
the high energy radiation data.   
For the 24 GeV Si data $\alpha$ is consistent with zero within errors, whereas
for the diamond data $\alpha$ is found negligible. 
Table~\ref{table:damage_constants} shows the results. The damage constant $k$ for diamond is generally smaller than for Si by a factor of 2 - 3.
%
\begin{table}[h]
\caption{Damage constants obtained for diamond and Si sensors irradiated by 25 MeV and 24 GeV protons, respectively.}
\begin{center}
\begin{tabular}[t]{|c|c|c|}
  \hline
  ~~                  	& 25 MeV protons            & 24 GeV protons                 \\
  \hline
  $k_{diamond}$		& $3.02^{+0.42}_{-0.36}$   & $0.69^{+0.14}_{-0.17}$        \\
  \hline
  $k_{Si}$              & $10.89^{+1.79}_{-1.79}$  & $1.60^{+0.38}_{-0.38}$        \\
  \hline
\end{tabular}
\end{center}
\label{table:damage_constants}
\end{table}

\begin{figure}
\begin{center}
\subfigure[25 MeV proton irradiation]{
    \includegraphics[width=0.45\textwidth]{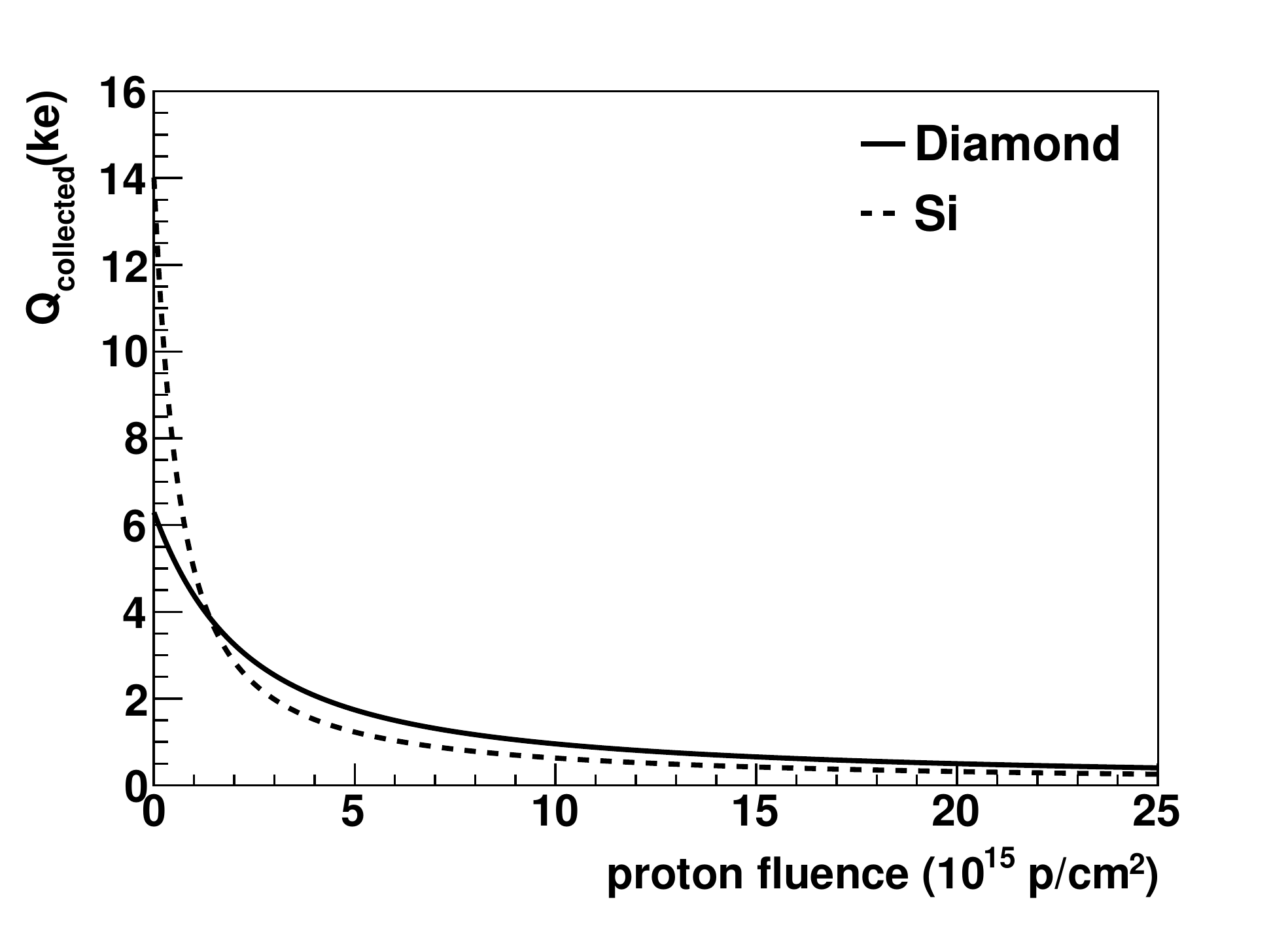}
    \label{fig:signal_charge_25MeVp}   }
    \hskip 0.5 cm
\subfigure[24 GeV proton irradiation]{
    \includegraphics[width=0.45\textwidth]{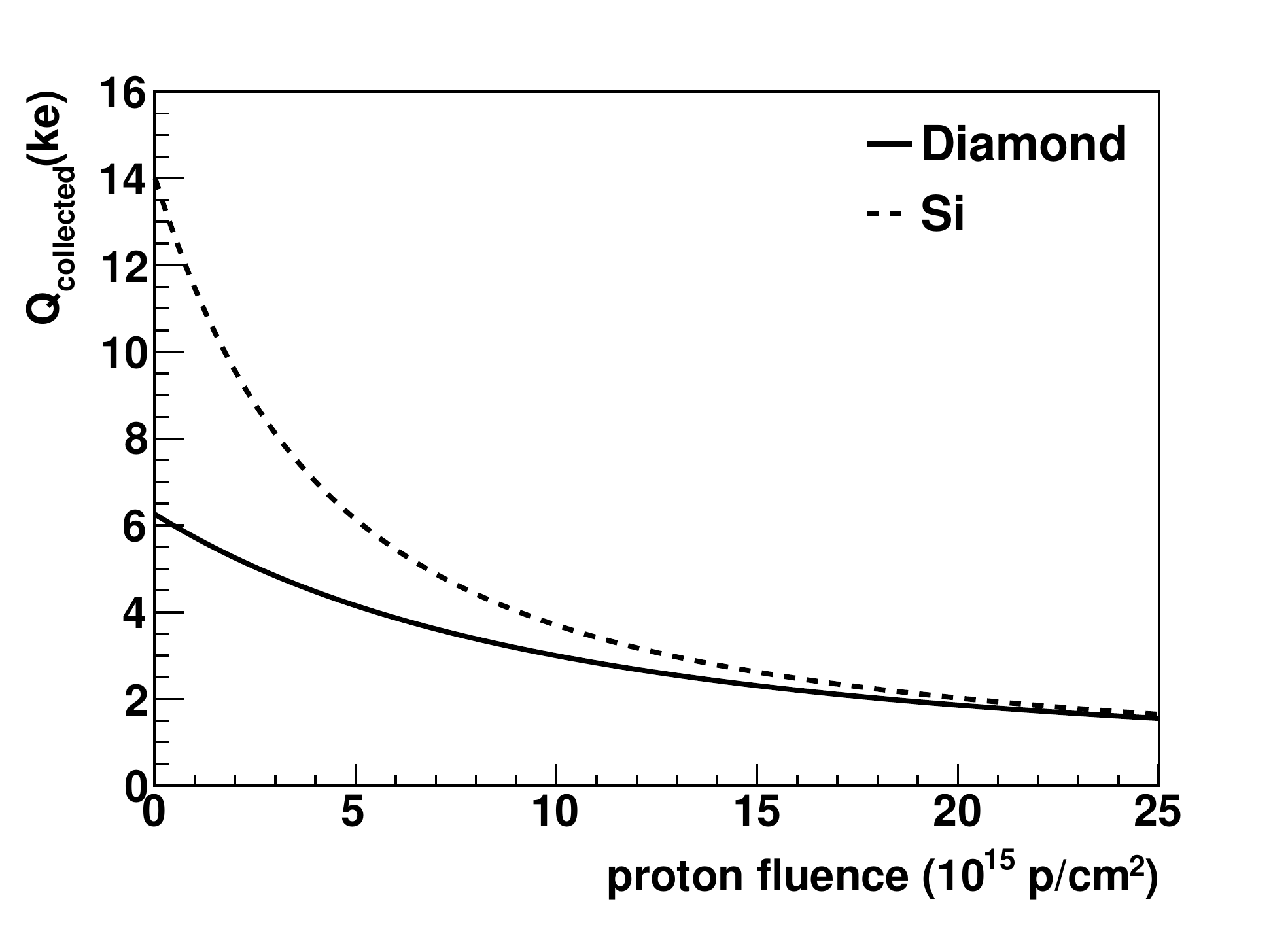}
    \label{fig:signal_charge_25GeVp}  }
\caption[]{Expected signal (MPV) of a minimum ionizing particle in units of 1000 electrons for 200 $\mu$m thick diamond (scCVD) and planar silicon sensors damaged by 25 MeV (a) and 24 GeV (b) proton irradiation. Charge multiplication is not considered here (see text eq.\eqref{eq:damage_curve_2}).\label{fig:signal_charge}}
\end{center}
\end{figure}

With the damage constants $k$, eq.~\eqref{eq:damage_curve}, and relation~\eqref{eq:mfp_to_CCD} we can determine 
the signal charge at any fluence value for a given sensor thickness. For a minimum ionizing particle ($\beta\gamma$ = 100) and for
200 $\mu$m thick sensors the signal charge (without charge multiplication)  is plotted in Fig.~\ref{fig:signal_charge} as a function of the fluence. While for high energy radiation (24 GeV protons) Si keeps a larger signal yield compared to diamond, for the more damaging low energy radiation (25 MeV protons) the signal yield for diamond becomes comparable or even a bit better than Si at fluences of about 2 $\times$ 10$^{15}$ cm$^{-2}$ and above.  
%
\section{The Noise}\label{sec:noise}
\begin{figure}[thb]
\begin{center}
\includegraphics[width=0.8\textwidth]{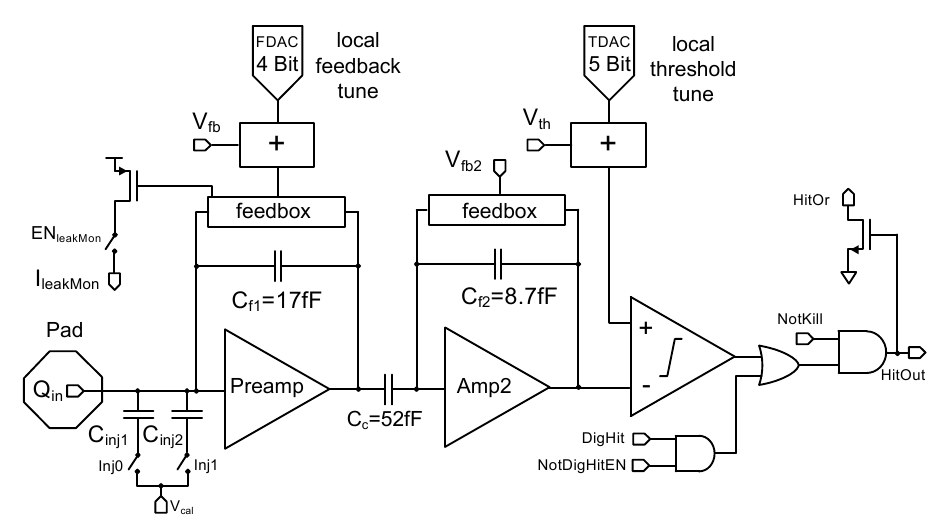}
\caption[]{Schematic diagram of the FE-I4 pixel cell readout circuitry.}
\label{fig:pix_circuit}
\end{center}
\end{figure}

The FE-I4 chip~\cite{FE-I4_MB2009,FE-I4_MGS2011} is a two-stage charge sensitive amplifier featuring constant current feedback in both stages
followed by a discrimination stage. The two amplifier stages are capacitively coupled. A leakage current compensation circuit
is implemented at the input to compensate sensor leakage current up to 100 nA. The chip has been optimized for a
maximum input capacitance of 400 fF. Figure~\ref{fig:pix_circuit} shows a block diagram of the FE-I4 cell circuitry.

For the noise estimate of a pixel sensor - readout chip configuration we employ
(a) a calculation using a noise model assuming some simplifications detailed below and (b) detailed noise simulations on the original
FE-I4 chip layout using the Virtuoso SPECTRE Circuit Simulator simulation package in CADENCE~\cite{SPECTRE}. 

\subsection{Calculation of FE-I4 noise}\label{sec:noise_calculation}
\begin{figure}[thb]
\begin{center}
\includegraphics[width=0.9\textwidth]{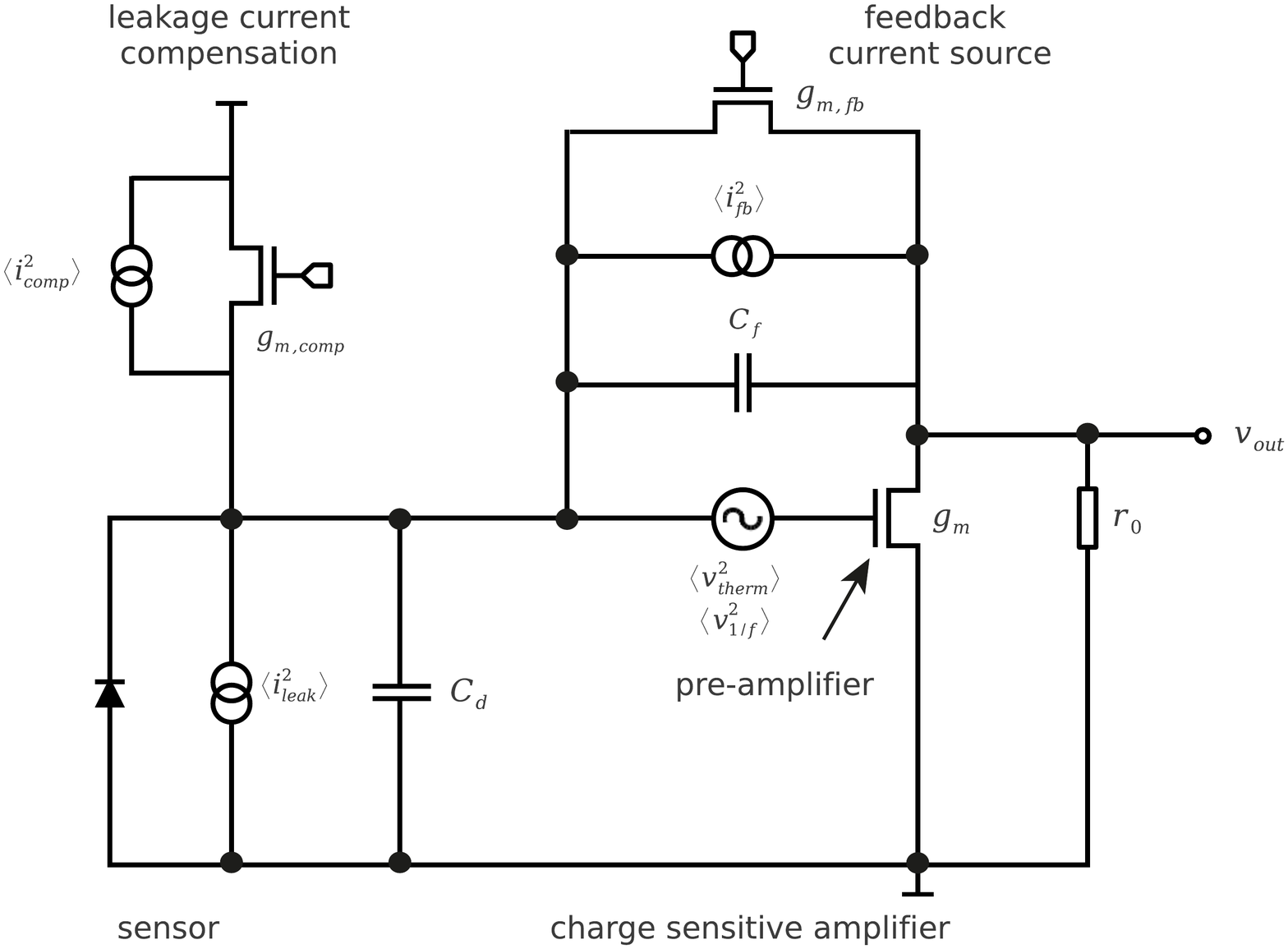}
\caption[]{Noise sources considered in the noise calculation (first stage only).}
\label{fig:pix_model}
\end{center}
\end{figure}

For the noise calculation the contributions of the second stage of the FE-I4 amplifier have been neglected and only the   
dominant noise sources shown in Fig.~\ref{fig:pix_model} have been considered:
thermal and 1/f-noise in the transistor channel which appear as series voltage noise at the input of the preamplifier; 
shot noise from the sensor leakage current and thermal noise from the leakage current compensation transistor, as well as thermal noise in the feedback loop (see Fig.~\ref{fig:pix_model}) which constitute a parallel noise contribution. 

The noise sources mentioned above generate a noise voltage at the output of the amplifier. Integrating the 
total noise power spectral density multiplied by the amplifier transfer function we obtain the following formula for the equivalent noise charge:
\begin{eqnarray}\label{eq:ENC}
\langle ENC^2 \rangle & = &  
A_{parallel} \cdot \frac{\tau_b^2}{\tau_b+\tau_c} 
+ A_{serial}^{thermal} \cdot \left [ \frac{1}{\tau_c}+\frac{\tau_a^2}{2\tau_c^3} \right ] 
+ A_{serial}^{\frac{1}{f}} \cdot \left [ \frac{\tau_a^2}{2\tau_c^2} + \ln \left ( \frac{\tau_b}{\tau_c} \right ) \right ] \\
A_{parallel}  & = &  \frac{2 i_{leak} + i_{fb}}{2q} \\
A_{serial}^{thermal} & = &  \frac{C_D^2}{q^2} \frac{kT}{2g_m} \\
A_{serial}^{1/f} & = &  \frac{C_D^2}{q^2}\frac{K_F}{C_{ox}WL} 
\end{eqnarray}

Here $\tau_{a} = C_{f}/g_m$, $\tau_{b} = C_{f}/g_{ds,fb}$, $\tau_{c} = C_{D}/g_{m}$ are time constants.
The drain output conductance of the feedback transistor $g_{ds,fb}$ is feedback current dependent. A typical value for $i_{fb}$ for  signals just above threshold is 4.5 nA and $g_{ds,fb} = 800$ nS.
$C_{f}$ is the feedback capacitance of the first stage CSA.
$i_{leak}$ is the leakage current.
$g_m$ = 225$\mu S$ is the transconductance of the preamplifier input transistor. 
The FE-I4 nominal bias current $i_{bias}$ is 7.5$\mu$A.
$K_{F} = 13.5 \times 10^{-25}$J is the 1/f noise coefficient.
$C_{ox}$, $W$, and $L$ are the oxide capacitance, width, and gate length of the MOSFET transistor.

Note that the noise performance depends to a large extent on the detector leakage current ($i_{leak}$) and the value of the
input capacitances to the preamplifier $C_D$, both of which need to be determined.
The detector leakage current, $i_{leak}$, is obtained for a given fluence as described by eq.~\eqref{eq:ileak_vs_fluence} below.
The precise values of pixel sensor capacitances $C_D$ are a priori not well known and have always been an uncertainty
in the characterization of pixel performance. We therefore have measured $C_D$  
as described in section~\ref{sec:pixcap_measurements} using a dedicated capacitance measurement chip.

While diamond - owing to its large band gap - can be assumed to have negligible leakage current even after strong irradiation, the leakage current change in silicon due to irradiation follows the relation (NIEL approximation)~\cite{Lindstroem}:
\begin{eqnarray}\label{eq:ileak_vs_fluence}
i_{leak } & = & i_{leak,0} + \alpha_i  \, \Phi \cdot V
\end{eqnarray}
where V is the volume under the pixel electrode, $\alpha_i$ is the damage factor for silicon ($\alpha = 4.0 \times 10^{-17} \frac{A}{cm}$~\cite{Lindstroem}) and hence $i_{leak}$ can be calculated.
The predicted NIEL linear behavior has been verified for silicon by measurements up to fluences of 10$^{15} n_{eq}$/cm$^2$~\cite{Moll199987}. It should be noted that the measurements in~\cite{Moll199987} are obtained from large area diodes and should be considered optimistic what segmented Si detectors is concerned.
Even if assumed to be zero before irradiation, $i_{leak}$ can become fairly large ($\cal{O}$ (100 nA)) 
at large fluences in excess of 10$^{15} n_{eq}$/cm$^2$ as expected at the LHC-upgrades.
Furthermore, a temperature correction has been applied, because the typical operation temperature ($T$) at LHC is -10$^\circ$C while the measurements reported here have been done at room temperature ($T_{ref}$):
\begin{eqnarray}\label{eq:ileak_tempcor}
i_{leak}(T) & = & i_{leak}(T_{ref}) \cdot R(T) \nonumber \\
R(T) & = & \left ( \frac{T}{T_{ref}} \right )^2 \exp \left [ \frac{-E_g}{2k_B} \left ( \frac{1}{T} - \frac{1}{T_{ref}} \right )\right ] \   ,
\end{eqnarray}
where $E_g$ is the band gap of Si, and $k_B$ is the Boltzmann constant.

\subsection{Noise simulation}\label{sec:noise_simulation}
The noise of the FE-I4 based pixel detector was also determined by simulations of the analog part of the FE-I4 using the 
Virtuoso Spectre Circuit Simulator~\cite{SPECTRE}.

{\bf AC noise simulation} is a common approach for small signal noise analysis which employs a linearized model of the circuit 
at its DC operating point. The noise is simulated in the frequency domain, treating the signal and noise sources separately assuming that they do not influence each other.

The {\bf transient noise simulation} is performed in the time domain. It is better suited than the AC simulation 
for our noise studies which evaluate the noise by S-curve measurements using multiple charge injections. Fluctuations of the signal peak when 
reaching the applied threshold voltage are of interest. In this case the
operation point is far away from its stable DC value and hence the accuracy of the AC simulation is limited. 
In the transient noise simulation, each component of the electronic circuit contains a noise source generating random noise signals in the time domain with the appropriate power spectral density distribution. The circuit voltages and currents are computed solving differential equations by effective Monte Carlo methods.  This method allows fairly realistic noise studies of non-linear and non-time invariant systems. In the simulation, a noise frequency band from 10 kHz to 1 GHz was chosen to extend well beyond the circuit bandwidth.
  
\begin{figure}
\begin{center}
\subfigure[ENC versus detector capacitance]{
	\includegraphics[width=0.45\textwidth]{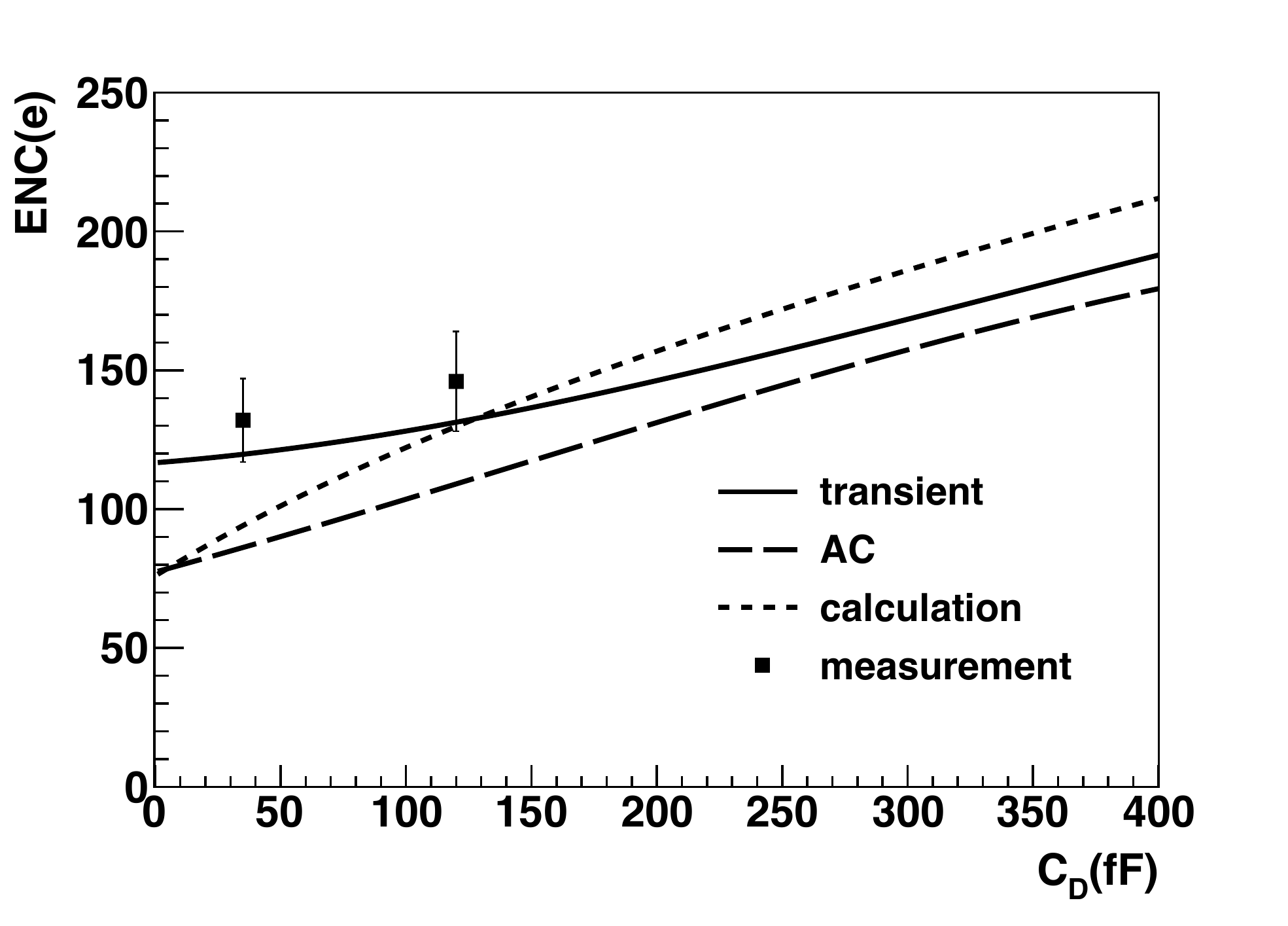}
	\label{fig:ENC_vs_Cd}
	}
	 \hskip 0.5 cm
\subfigure[ENC versus fluence ]{
	\includegraphics[width=0.45\textwidth]{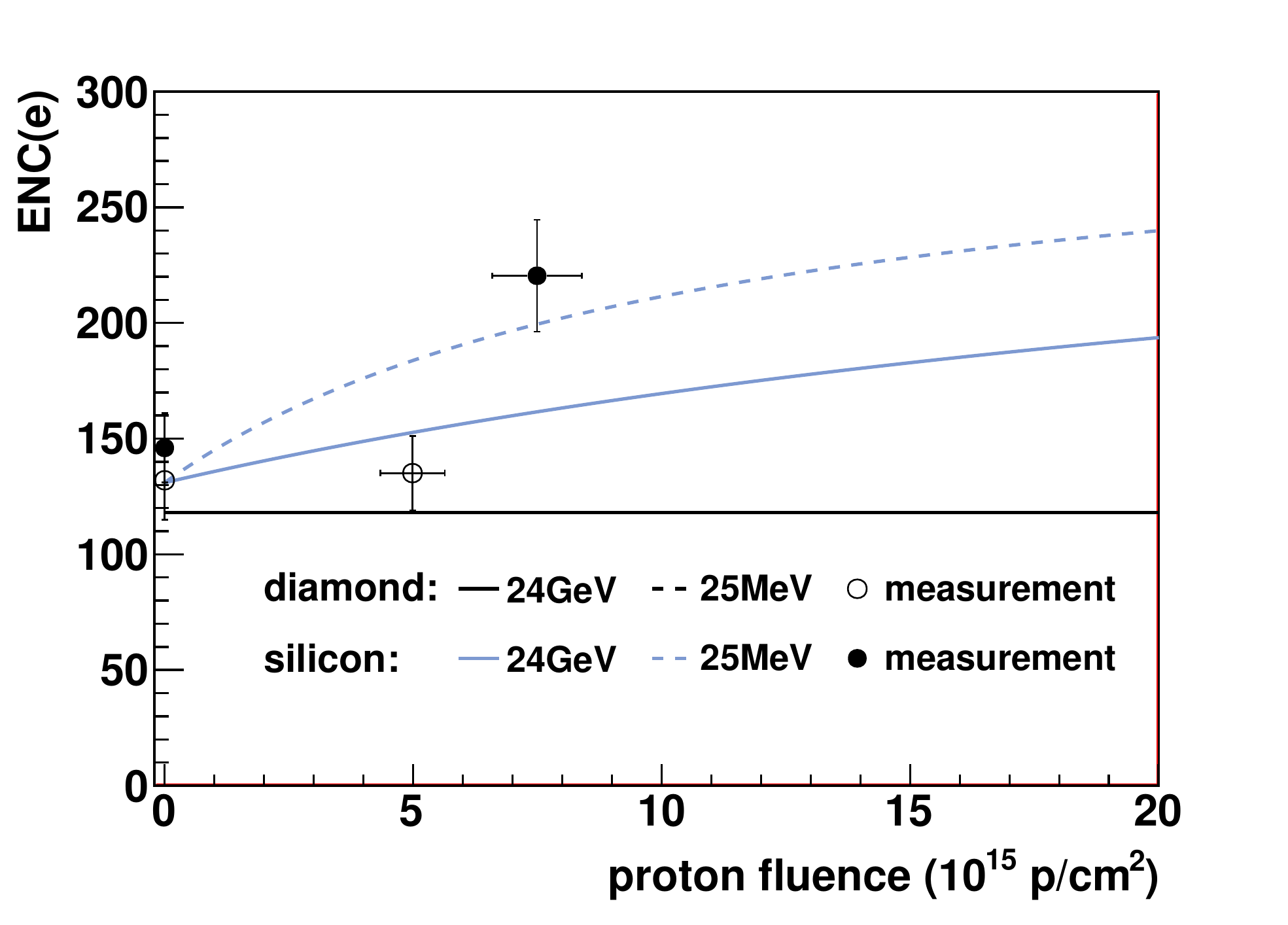}
	\label{fig:ENC_vs_ileak}
	}
\caption[]{Equivalent noise charge (ENC) as a function of two important noise parameters: (a) the detector capacitance $C_D$ and
(b) the leakage current $i_{leak}$. In (a) three noise estimation methods (see text) are compared with each other and with measurements (full squares): analytical calculation (dotted line), AC noise simulation (dashed) and transient noise simulation (solid). 
In (b) the best method (transient noise simulation) is shown for diamond ($C_D$ = 35 fF, dashed) and for silicon ($C_D$ = 120 fF, solid) as a function of the radiation fluence for 25 MeV and 24 GeV protons, respectively. The measurements are obtained from 
diamond (open circles) and silicon (full circles) devices irradiated by 25 MeV protons.

	} \label{fig:ENC_comparisons}
\end{center}
\end{figure}

In Fig.~\ref{fig:ENC_comparisons} we show the dependence of the noise calculations and simulations as a function of the most important parameters in this estimate, $i_{leak}$ and $C_D$. 
For Fig.~\ref{fig:ENC_vs_Cd}  $i_{leak} = 0 $ has been assumed. 
For Fig.~\ref{fig:ENC_vs_ileak} pixel capacitances of $C_D = 120$ fF have been assumed for silicon, 
while for diamond $C_D = 35 fF$ is assumed (see section~\ref{sec:pixcap_measurements}).
Noise measurements obtained with (a) unirradiated devices for diamond and silicon and (b) with silicon pixel assemblies
irradiated to a fluence of 10$^{15}$ n$_{eq}$/cm$^{-2}$ corresponding to $i_{leak} = 110 nA$ (eq.~\ref{eq:ileak_vs_fluence}) 
are included in Fig.~\ref{fig:ENC_comparisons} to verify the noise predictions. The agreement is remarkably good.

Given that the calculation and the transient noise simulation agree at the 5$\%$- and 20$\%$-level, and also agree reasonably well with measurements of the noise, we assume the transient noise model for the following SNR estimations.

\subsection{Measurement of $C_D$ and $C_{in}$}\label{sec:pixcap_measurements}
The knowledge of the total input capacitance coupled to the preamplifier is an important ingredient for a reliable comparison between diamond and silicon. It receives contributions from the sensor's interpixel capacitance (dominant), its capacitance to the backside, the capacitances of the solder bump and the bump pad, as well as the amplifier input capacitance.
While in the case of diamond the interpixel capacitance is geometrically defined by the pixel metallization, for silicon the pixel implant geometry and the (punch through) biasing grid geometry define the pixel capacitance. Thus a simple scaling with the dielectric constants of both materials is not expected.
In order to obtain reliable values of $C_D$ for diamond and silicon pixels with FE-I4 pixel sizes and pitch, we have developed a capacitance measurement chip (PixCap)~\cite{pixcap} for the measurement of the capacitance with high accuracy. The chip 
has been bonded to the pixel sensors using the bump and flip-chip technology. 

\begin{figure}
\begin{center}
\subfigure[PixCap principle]{
	\includegraphics[width=0.5\textwidth]{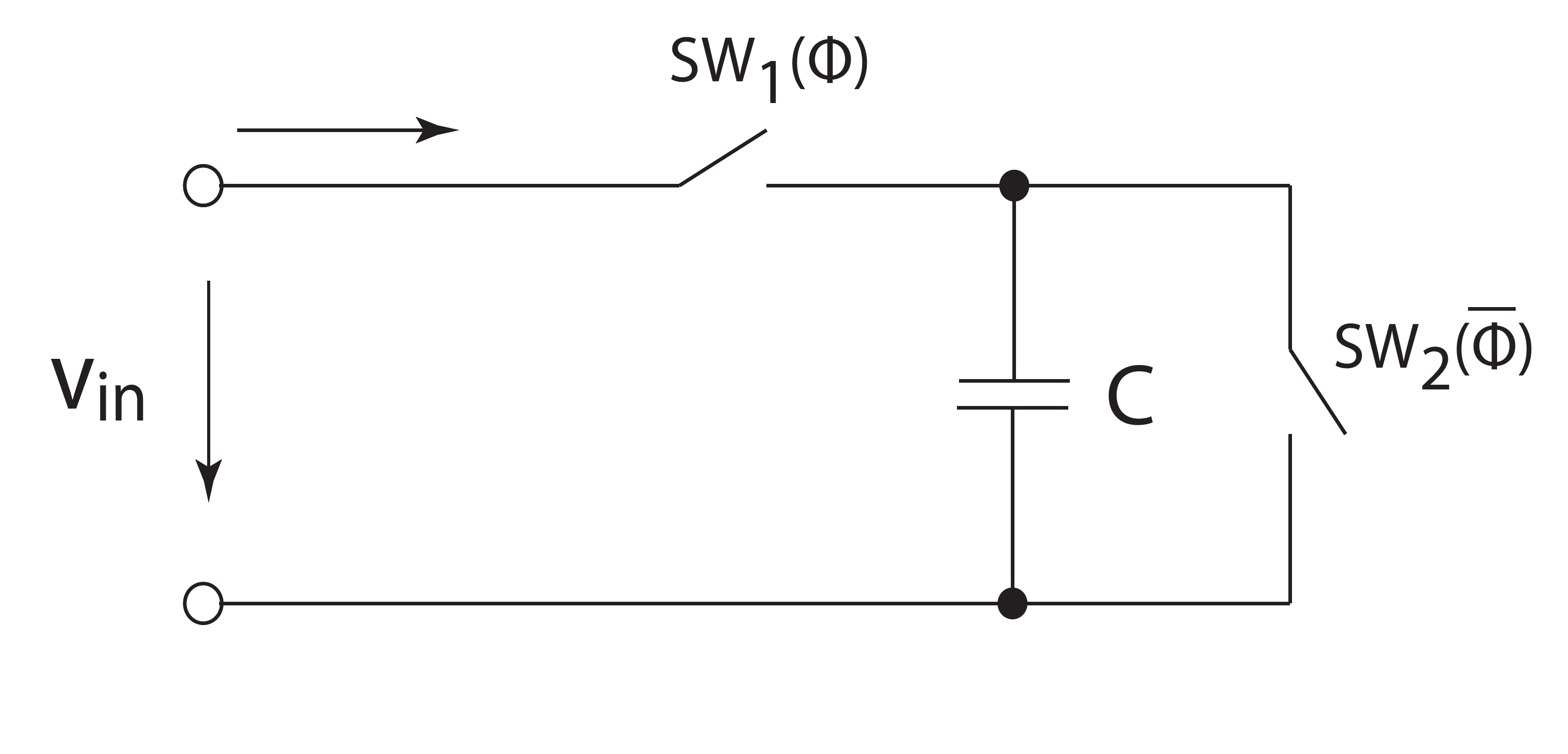}
	\label{fig:pixcap_circuit}
	}
	 \hskip 0.5 cm
\subfigure[PixCap chip bonded to sensor]{
	\includegraphics[width=0.3\textwidth]{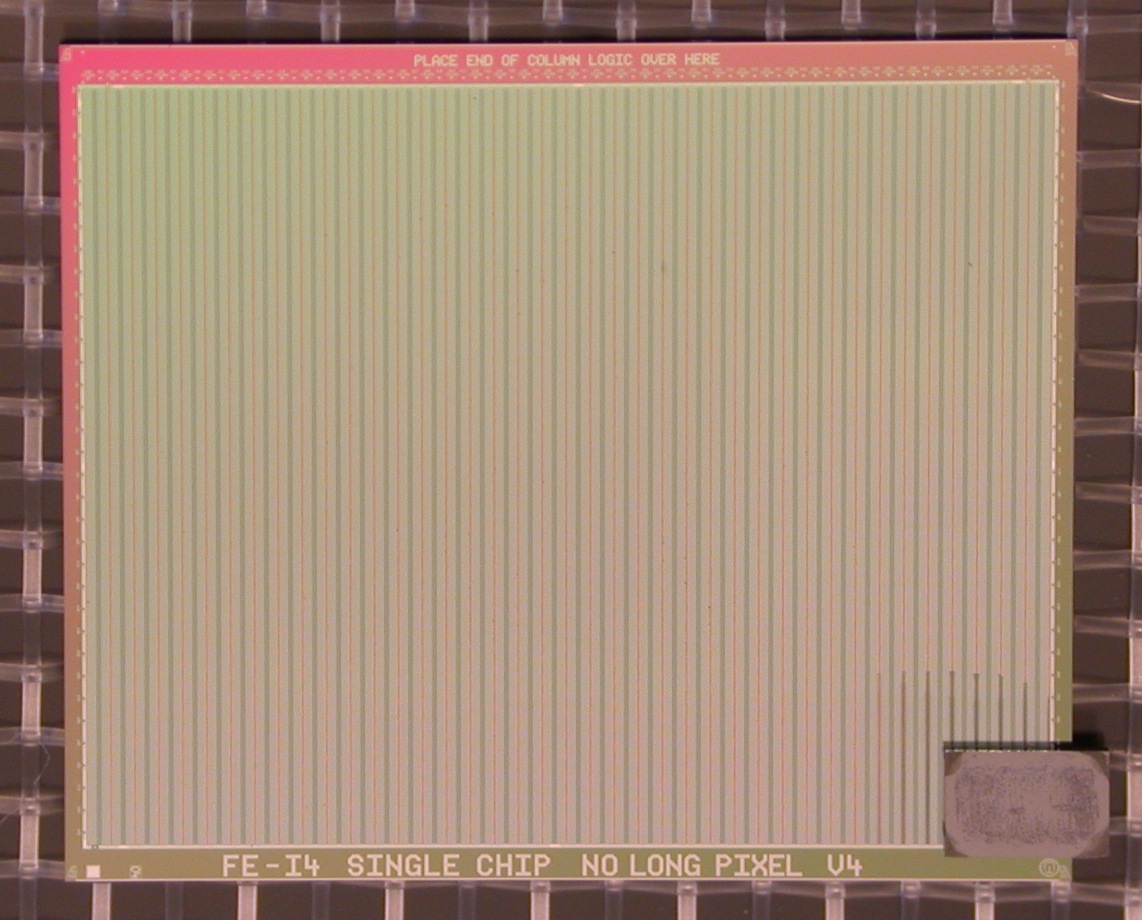}
	\label{fig:pixcap_photo}
	}
\caption[]{Charge pump principle (a) and photo (b) of the PixCap chip bonded to a pixel sensor (here planar silicon) .
For economic reasons the PixCap chip is much smaller in area than the sensor covering only $40 \times 8$ pixels.
	} \label{fig:pixcap}
\end{center}
\end{figure}

The measurement principle is based on a switching circuit (Fig.~\ref{fig:pixcap_circuit}) that charges and discharges the capacitance to be measured 
with a chosen frequency $f$. By measuring the average switching current at a well determined frequency and input voltage the capacitance is extracted:  
\begin{eqnarray}\label{eq:pixcap_measurement}
C & = & \frac{Q}{V_{in}} = \frac{\int i(t) dt}{V_{in}} = \frac{\langle I \rangle - \langle i_{leak} \rangle}{V_{in} \times f}
\end{eqnarray}
where $V_{in}$ is the input voltage and $f$ the switching frequency (here 4 MHz).
Leakage current is corrected for by dedicated measurements~\cite{pixcap}.
\begin{figure}
\begin{center}
\subfigure[diamond pixels]{
	\includegraphics[width=0.4\textwidth]{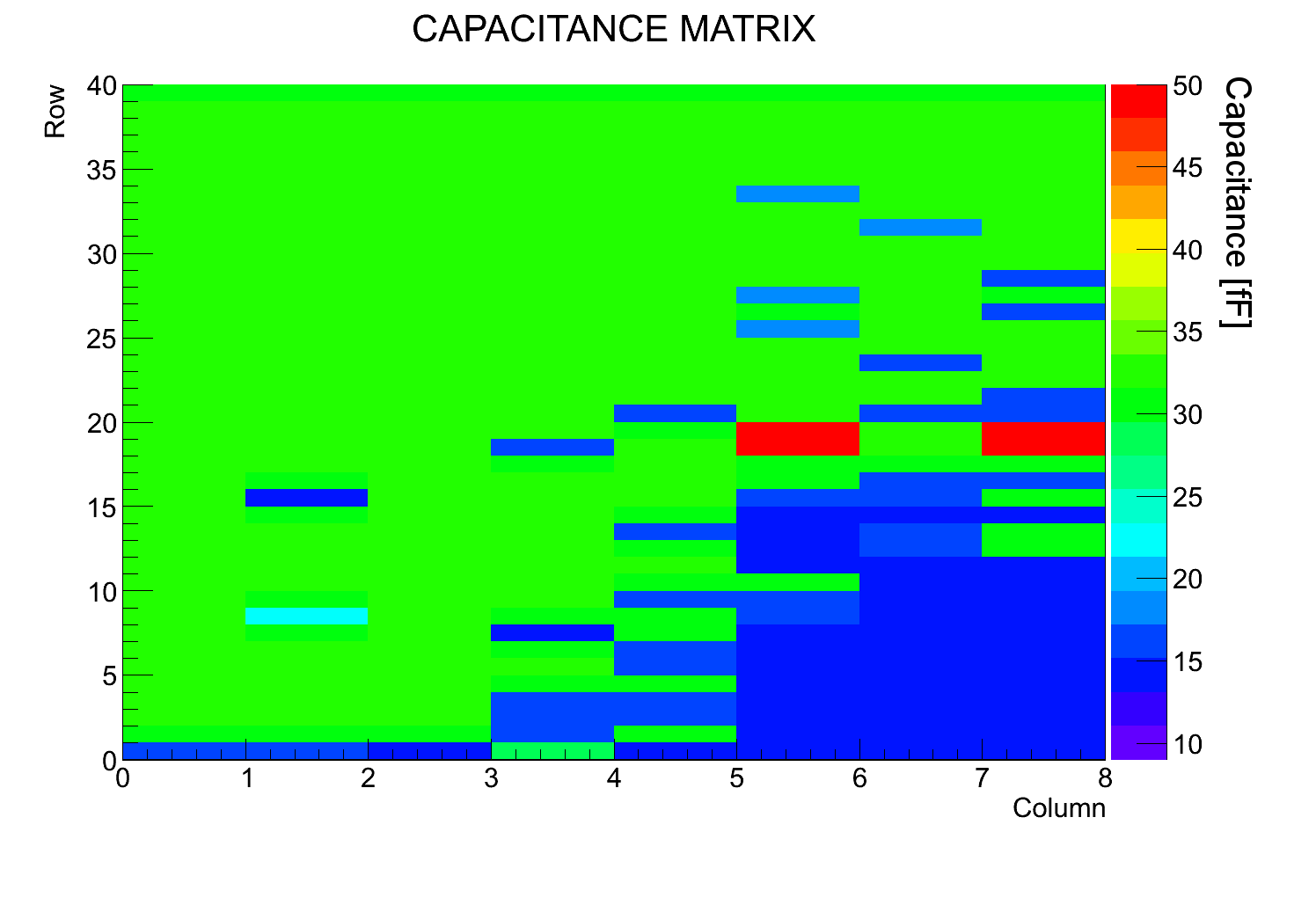}
	\label{fig:pixcap_map_diamond}
	}
	 \hskip 0.5 cm
\subfigure[planar silicon pixels]{
	\includegraphics[width=0.4\textwidth]{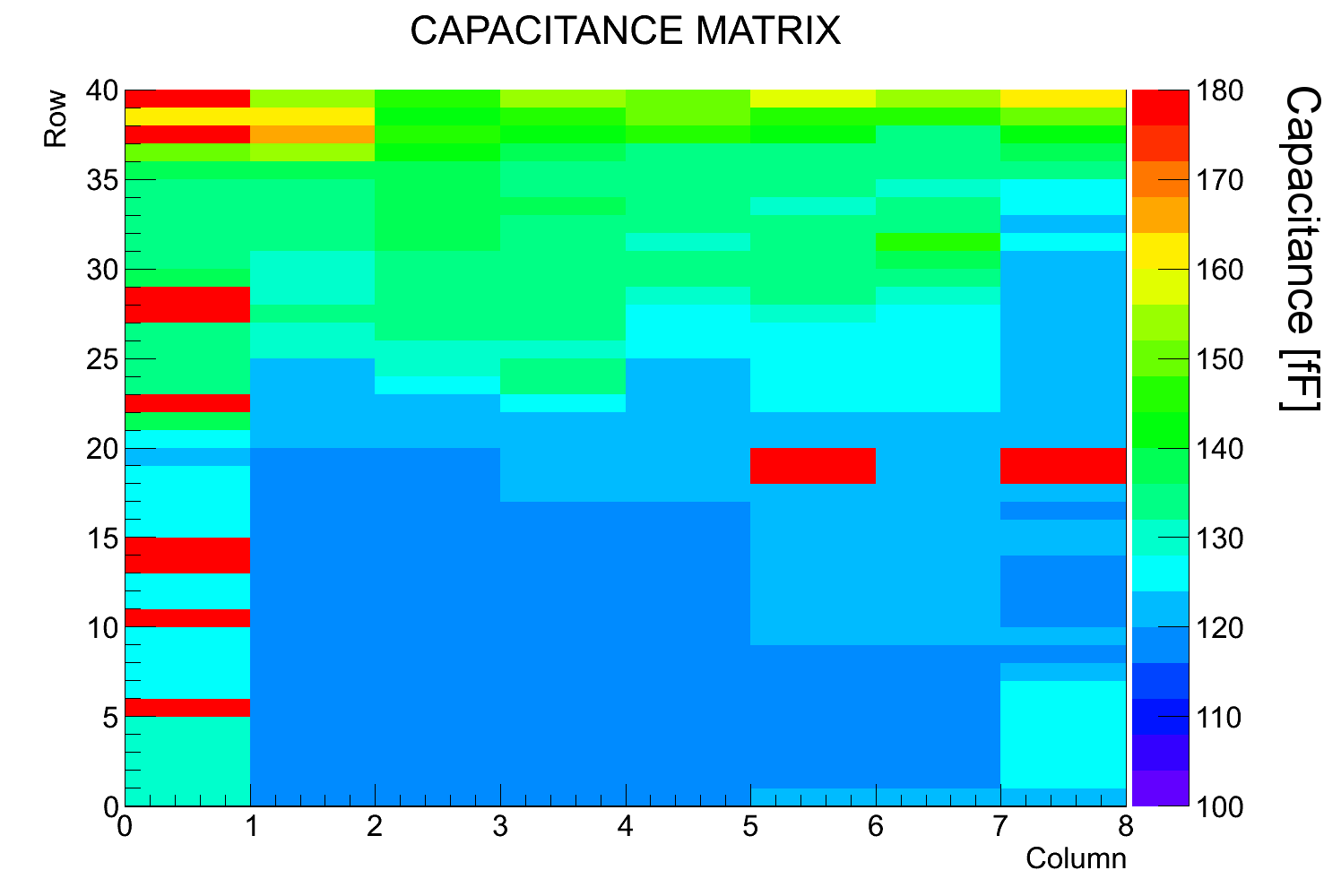}
	\label{fig:pixcap_map_silicon}
	}
\caption[]{Capacitance maps of (a) a 450 $\mu$m thick diamond pixel sensor (60V) and (b) a 250 $\mu$m thick planar silicon pixel sensor (80V, fully depleted) measured with the PixCap chip. The sensors are unirradiated. The two pixels marked red are test pixels which have additional capacitors connected to their inputs.
Note that color axis is zero suppressed.
The area with low capacitance measurements in (a) is due to bonding problems. \label{fig:pixcap_maps}    }
\end{center}
\end{figure}

By individually programming the switching sequences for neighboring pixels the
different contributions to the total capacitance can be individually determined. The measurement error depends on the
leakage current and the capacitance value and lies around 0.1 fF  for diamond and 0.3 fF for silicon. 
The capacitance measurements yield the following results.  Without any sensor
bonded, the input capacitance to the FE-I4 preamplifier is $C_{no\, det} = (11.6 \pm 0.1) fF$.  The capacitance of the 
input transistor transistor itself is $C_{transistor} = (32 \pm 2.9) fF$ . 
Bonded to a diamond sensor the map of measured pixel capacitances
is shown in Fig.~\ref{fig:pixcap_map_diamond}. The sensor was biased with 60V, although no dependence of the bias voltage is observed. The average pixel capacitance measured for diamond bumped to the FE-I4 cell, i.e. including all related connection capacitances but without the transistor capacitance $C_{transistor}$ is measured to be
$$
\langle C_D^{diamond} \rangle = (33 \pm 1) fF \ ,
$$
where the error includes the spread over the considered region of the pixel matrix (see Fig.~\ref{fig:pixcap_maps}).
For silicon there is a dependence on the applied bias voltage until the sensor is fully depleted and neighboring pixels are isolated. 
For unirradiated sensors this is the case for voltages above 60 V.  The measurements were done at 80 V. The measured capacitance map is shown in 
Fig.~\ref{fig:pixcap_map_silicon}. The average value for silicon is
\begin{eqnarray*}
\langle C_D^{Si} \rangle & =  & (117 \pm 2) fF  \qquad \mathrm{planar\ electrodes, sample\  A} \\
\langle C_D^{Si} \rangle & =  & (123 \pm 4) fF  \qquad \mathrm{planar\ electrodes, sample\  B} \\
\langle C_D^{Si} \rangle & = & (181  \pm 2) fF \qquad \mathrm{3D\ electrodes}
\end{eqnarray*}
The capacitance quoted here for 3D silicon pixels is for a geometry with two 3D electrodes under the pixel area of the FE-I4 chip.

\section{Signal to noise ratio at high particle fluence}\label{sec:SNR}
With the measured capacitances and the fluence dependent leakage current calculations, it is now possible to predict the noise as a function of the fluence for a given energy of the damaging radiation.
In order to compare diamond in a reasonable way with silicon sensors, some assumptions (for silicon) have been made in this 
paper which are summarized here again:
\begin{itemize}
\item[-] The electrode geometry changes the characteristics of a silicon pixel detector. Currently only two viable options exist, sensors with planar pixel electrodes
on one side of the sensor~\cite{planar-pixels} and sensors with cylindrical electrodes etched into the silicon bulk, so called 3D-Si-sensors~\cite{3D-Si}. We chose here
to base our comparison with diamond on planar pixel sensors mated to the identical FE-I4 pixel chip.
\item[-] In contrast to diamond, silicon needs to be made free of charge carriers by depletion. For increasing irradiation the voltage which is needed for full depletion of a Si-sensor needs to be increased.
This will be possible only to a certain maximal extent in a large area pixel detector system and will be limited by the power that can be afforded.
At fluences much above 10$^{15}$ n$_{eq}$ cm$^{-2}$, a 200 $\mu$m thick Si-sensor can usually no longer be fully depleted, and the depletion depth
will change with increasing irradiation even if the bias voltage is steadily increased.

We hence base our comparison on irradiation measurements made by Affolder et al.~\cite{Affolder_25MeV,Affolder_24GeV} for 310 $\mu$m thick silicon sensors and assume that the maximum bias voltage to be 600 V\footnote{Values for this bias setting have been obtained by averaging the measurements quoted in ref.~\cite{Affolder_25MeV,Affolder_24GeV} for 500 V and 700 V}
which corresponds to the current power supply limit in ATLAS. 
\item[-] In thin silicon sensors irradiated to fluences up to 5.6$\times$ 10$^{15}$ n$_{eq}$ cm$^{-2}$, charge multiplication has been observed~\cite{Casse:2010zz}. We ignore this
possibility for this study, because it has not been demonstrated yet whether charge multiplication can be exploited 
for a (homogeneous) operation of pixel detectors for
particle detection after large fluences. It is also possible that similar effects can occur in diamond detectors in high electric fields.
\end{itemize}
We conclude therefore, that our assessment of diamond pixels compared to planar Si-pixel detector assemblies,
both bonded to the FE-I4 pixel readout chip, constitutes a reasonable and fair choice for a comparison, notwithstanding the possibility that different operation points and designs may partly impart our conclusions.

\begin{figure}
\begin{center}
\subfigure[SNR for 200 $\mu$m sensor thickness]{
	\includegraphics[width=0.45\textwidth]{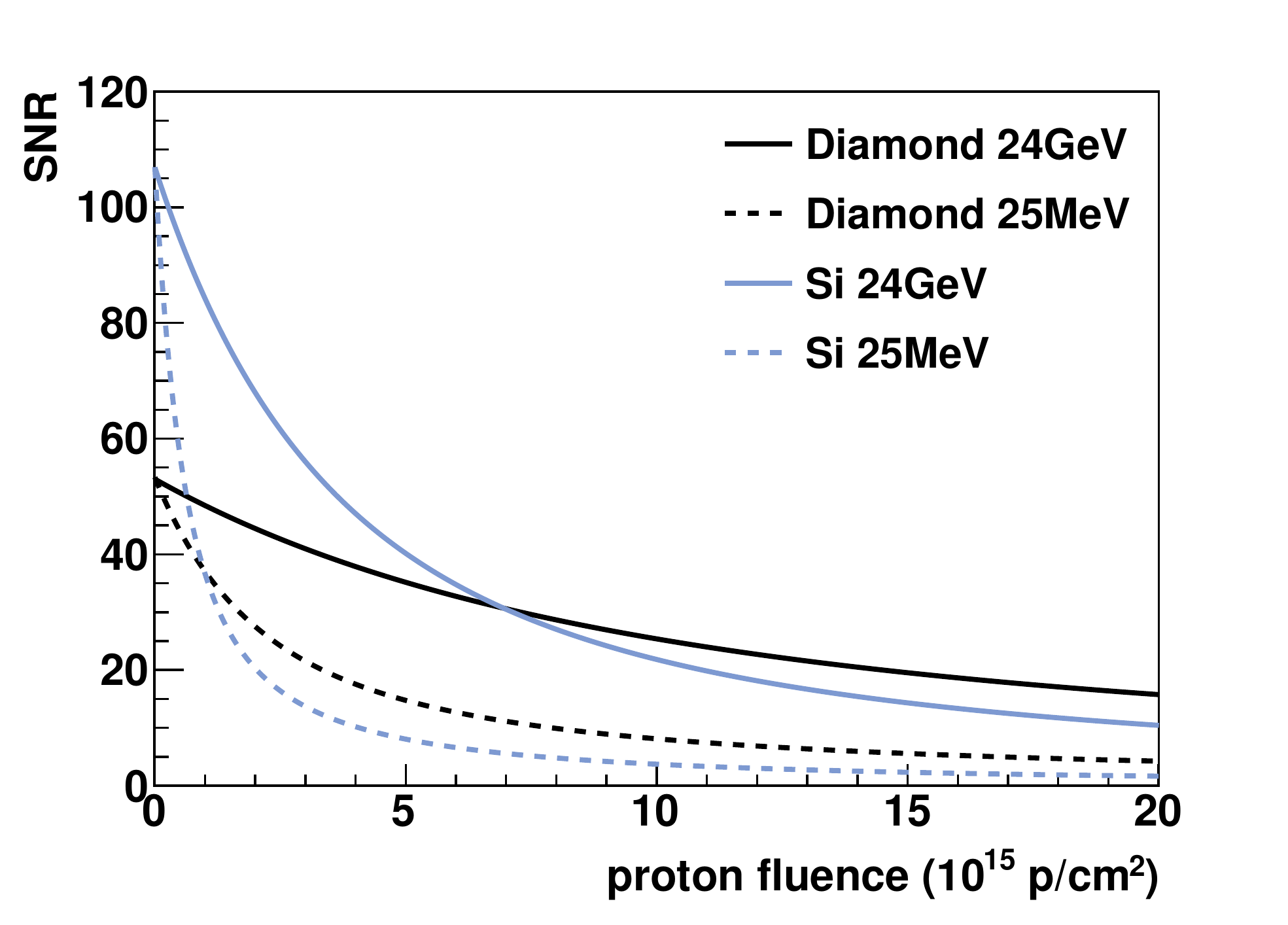} \label{fig:SNR_200}
	}
	 \hskip 0.5 cm
\subfigure[SNR for sensors with 0.1$\%$ x/X$_0$]{
	\includegraphics[width=0.45\textwidth]{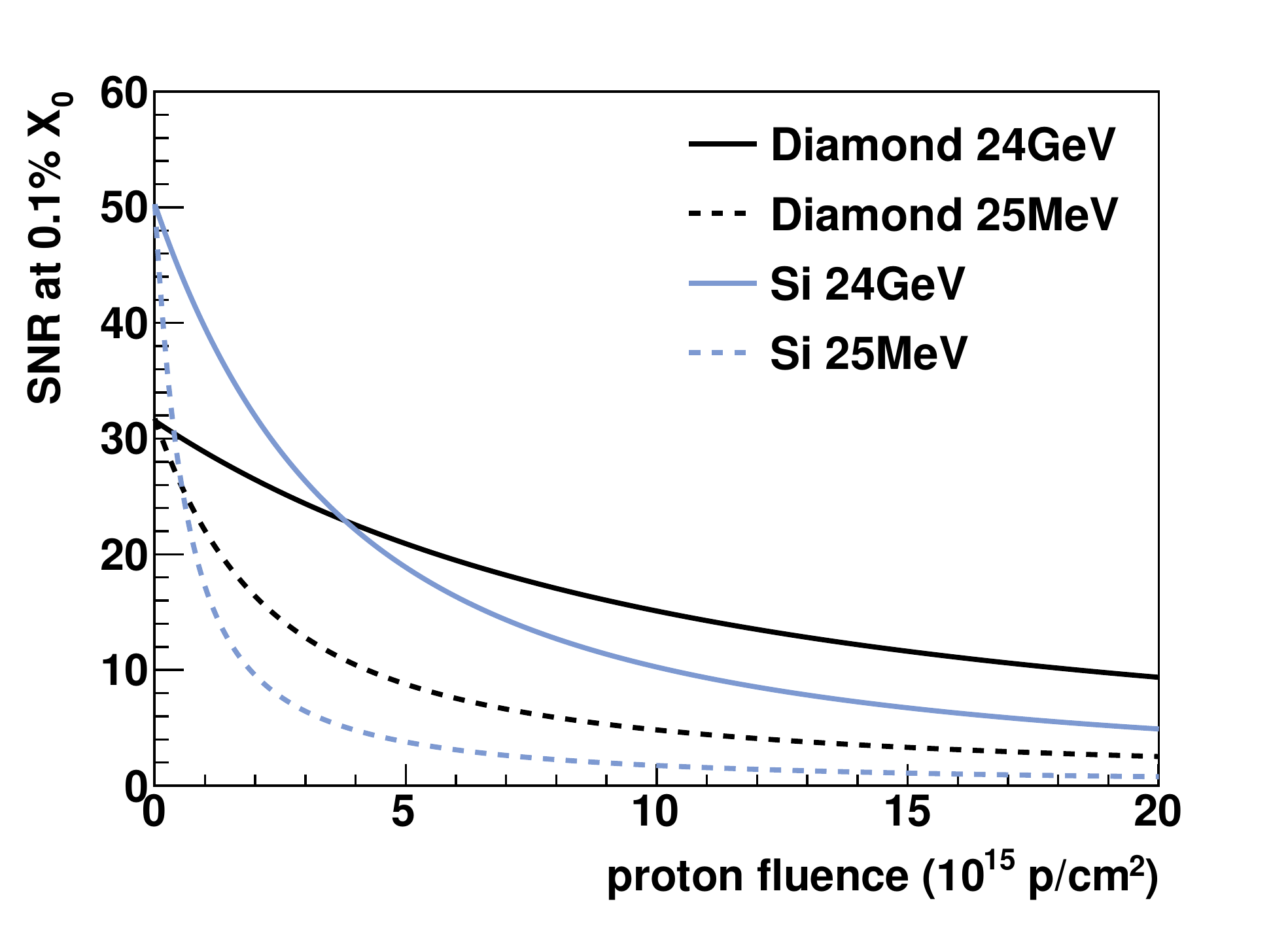} \label{fig:SNR_X0}
	}
\caption[]{Signal-to-noise ratio (SNR) as a function of proton radiation fluence for (a) sensors of an assumed  thickness of 200 $\mu$m and (b) for diamond and silicon
sensors of the same thickness in units of radiation length (here $0.1 \% X_0$). Results for two different energies of the damaging radiation are shown. Diamond is shown as black lines, silicon as gray lines; dashes lines are for 25 MeV, solid lines for 24 GeV proton 
irradiation. Note that the results for diamond shown are for scCVD; for pCVD diamond the curves are shifted to the right by an amount given in the text.} \label{fig:SNR}
\end{center}
\end{figure}

With the measured signal development as a function of fluence from section~\ref{sec:signal} we plot the signal-to-noise ratio (SNR) for an assumed pixel sensor thickness of 200 $\mu$m and for the two different damage energies 
in Fig.~\ref{fig:SNR_200}. While the different responses to high and low energy irradiation lead to different SNR predictions the conclusion of Fig.~\ref{fig:SNR_200} is nevertheless evident:
diamond pixel modules exceed the performance of (planar) Si pixels in terms of the SNR at fluences above 
$10^{15}p/cm^{2}$, for scCVD between about $1 \times 10^{15}p/cm^{2}$ (25 MeV protons) and  $7 \times 10^{15}p/cm^{2}$ (24 GeV protons). For pCVD diamond this cross over point would be shifted to the right by $+3.8 \times 10^{15}$ p/cm$^{2}$ (24 GeV protons) and by $+ (1\ \rm{to}\ 2) \times 10^{15}$ p/cm$^{2}$ (25 MeV protons, see Fig.~\ref{fig:damage_curves}).

Since the question of how much material a pixel vertex detector or tracker constitutes is important in design
considerations for large collider detectors, a fair comparison should also account for this effect.
Fig.~\ref{fig:SNR_X0} therefore shows the same dependence as Fig.~\ref{fig:SNR_200}, but for sensors
of the same thickness in terms of radiation length: 0.1$\%$ X$_0$.  This normalization moves the
turn-over point in the diamond - silicon comparison to lower values of the radiation fluence, namely to about
$0.5 \times 10^{15}$ p cm$^{-2}$ (25 MeV protons) and $3.5 \times 10^{15}$ p cm$^{-2}$ (24 GeV protons), respectively. 
This comparison does not take into account that cooling and other services also play an important role in the material budget, 
in particular for silicon sensors which at HL-LHC require operation temperatures well below 0$^o$ 
C~\cite{ATLAS_pixel_paper_2008,CMS_pixel_paper_2008}. 

\section{Conclusions}\label{sec:conclusions}
This paper presents an assessment of the performance of diamond pixel detectors relative to silicon pixels at
very high radiation fluences quantified in terms of the signal-to-noise ratio (SNR). For the study pixel detector assemblies (single chip modules)
based on the ATLAS pixel readout chip FE-I4 have been used, both for diamond and for Si.
The individual ingredients to
this quantity are each determined whenever possible by measurements, aided by calculations/simulations as a function of fluence.
The signal development with fluence is extracted from dedicated irradiation campaigns up to fluences of 10$^{16}$ p/cm$^{2}$ at two different energies, 25 MeV and 24 GeV protons, as well as from published data on silicon~\cite{Affolder_25MeV, Affolder_24GeV}.
In order to properly treat the two materials with a large mobility-lifetime product, the mean free path $\lambda_{e/h}$ rather than the
charge collection distance $CCD$ has been employed. The damage constants determined from a fit to the damage curves are
2 - 3 times smaller for diamond than they are for silicon. Poly-crystalline diamond differs from mono-crystalline one in this 
by a shift of the damage curves to lower fluences by an amount of between 1 to 3.8 x 10$^{15}$ p/cm$^{^2}$, depending on the 
irradiation energy.
For the noise determination the actual layout of the FE-I4 pixel chip has been used in a transient noise simulation, backed by an analytical calculation.
The two most important parameters in the noise determinations, the leakage current $i_{leak}$ and the pixel input capacitance to the preamplifier $C_D$, have been either derived ($i_{leak}$) as a function of fluence using the NIEL assumption, or  measured ($C_D$) using a dedicated capacitance measurement chip (PixCap~\cite{pixcap}). 
The resulting SNR leads to the conclusion that the SNR of diamond pixel sensors exceeds that of planar Si at fluences above  10$^{15}$ p/cm$^{^2}$. 

\acknowledgments
We are grateful to the Fraunhofer FhG institute IZM, Berlin, especially T. Fritzsch, for their careful assembly of the various sensors.
and to Walter Ockenfels and Wolfgang Dietsche for the corresponding final assembly and wire bonding of the modules.
We thank Marlon Barbero and the FE-I4 team for advice. 
We would also like to thank the Irradiation Center at KIT, Karlsruhe and CERN where the irradiation of the devices 
used in this paper have been carried out, especially A. Dierlamm of KIT.
This work is supported by the Initiative and Networking Fund of the Helmholtz Association, contract HA-101 ("Physics at the Terascale"), by the German Ministerium f{\"u}r Bildung, Wissenschaft, Forschung und Technologie (BMBF) under contract no. 05H09PD2 and by the US Department of Energy grant DE-FG02-91ER40690.

    	\bibliographystyle{unsrt}
	\bibliography{Diamond_SNR_paper}

\end{document}